\documentclass[twocolumn,english,letterpaper,showpacs,floatfix]{revtex4}
\begin{document}
\newcommand{\be}{\begin{equation}}
\newcommand{\ee}{\end{equation}}
\newcommand{\bear}{\begin{eqnarray}}
\newcommand{\eear}{\end{eqnarray}}
\title{ A plausible upper limit on the number of e-foldings}
\author{Bin Wang}
\email{binwang@fma.if.usp.br}
\affiliation{ Department of Physics, Fudan University, Shanghai 200433,
P. R. China }
\author{Elcio Abdalla}
\email{eabdalla@fma.if.usp.br}

\affiliation{Instituto De Fisica, Universidade De Sao Paulo, C.P.66.318, CEP
05315-970,  Sao Paulo, Brazil }

\begin{abstract}
Based solely on the arguments relating Friedmann equation and the Cardy
formula we derive a bound for the number of e-folds during inflation for
a standard Friedmann-Robertson-Walker as well as non-standard four
dimensional cosmology induced by a Randall-Sundrum-type model.
\end{abstract}

\maketitle
Motivated by the well-known example of black hole entropy, an influential
holographic principle has put forward,  suggesting that microscopic
degrees of freedom that build up the gravitational dynamics actually
reside on the boundary of space-time \cite{thooft}. This principle
developed to the Maldacena's conjecture on AdS/CFT correspondence
\cite{maldacenaconj} and further very important consequences, such as
Witten's \cite{wittenidentification}
identification of the entropy, energy and temperature of CFT at 
high temperatures with the entropy, mass and Hawking temperature of the
AdS black hole \cite{hawpage}.

On the other hand, although standard model
of particle physics has been established as the uncontested theory of all
interactions down to distances of $10^{-17}$m, there are good reasons to
believe that there is a new physics arising soon at the experimental
level, and this fact is going to appear in a clear fashion in the setup of
cosmology, where a large number of quite exciting developments is giving
rise to a precision cosmology. Thus cosmology may provide an alternative
laboratory for string theory. From the 
eighties several authors tried to analyse the kind of cosmology arising
from string inspired models, which are essentially general relativity in
higher dimensions together with scalar and tensor fields. In case we
introduce also the brane concept, a consistent picture of the brane
universe is achieved, and we can describe the evolution of the universe by
means of solutions of the Einstein field equations in higher dimensions,
with a four dimensional membrane. We thus seek at a description of the
powerful holographic principle in  cosmological settings, where its
testing is subtle, and the question of hologrphy therein has been
considered by several authors \cite{severalauthors} have shown that for
flat and open Friedmann-Lemaitre-Robertson-Walker(FLRW) universes the area
of the particle horizon should bound the entropy on the backward-looking 
light cone. In addition to the 
study of holography in homogeneous cosmologies, attempts to generalize the
holographic principle to a generic realistic inhomogeneous cosmological
setting were carried out in \cite{tavaellis}. 
Later, the very interesting study of the holographic principle in FLRW
universe filled with CFT with a dual AdS description has been done by Verlinde
\cite{everlindeunpu}, revealing that 
when a universe-sized black hole can be formed, an interesting and surprising
correspondence appears between entropy of CFT and Friedmann equation
governing the radiation dominated closed FLRW universes. Generalizing
Verlinde's discussion to a 
broader class of universes including a cosmological constant
\cite{wangabdallasu}, we found that matching 
of Friedmann equation to Cardy formula holds for de Sitter closed and AdS flat
universes. However for the remaining de Sitter and AdS universes, the
argument fails due to breaking 
down of the general philosophy of the holographic principle. In high
dimensions, various 
other aspects of Verlinde's proposal have also been investigated in a
number of works \cite{all1}.

In a recent paper \cite{savoverli}, further light on the
correspondence between Friedmann equation and Cardy formula has been shed
from a Randall-Sundrum type brane-world perspective
\cite{randallsundrum}. Considering the CFT dominated universe as a
co-dimension one brane with fine-tuned tension in a background of an AdS
black hole, Savonije and Verlinde found the 
correspondence between Friedmann equation and Cardy formula for the
entropy of CFT when 
the brane crosses the black hole horizon. This result has been further
confirmed by studying a brane-universe filled with radiation and stiff-matter,
quantum-induced brane worlds and radially infalling brane
\cite{biswasmukh}. The discovered relation between Friedmann equation and
Cardy formula for the entropy shed significant light on the meaning  
of the holographic principle in a cosmological setting. However the
general proof for this correspondence for all CFTs is still difficult at
the moment. Other settings have also been considered as in
{\it e. g.} \cite{all2}. It is worthwhile to 
further check the validity of the correspondence in broader classes of
situations than \cite{everlindeunpu,savoverli}. 

Our motivation here is the use of the correspondence between the CFT gas
and the Friedmann equation and to establish an upper bound for the number of
e-foldings  during inflation, using a small number of assumptions. The
main point is an upper limit for entropy, a fact that we can derive from
the above correspondence. Recently, Banks and Fischler 
\cite{banksfischler} have considered
the problem of the number of e-foldings in a universe displaying an
asymptotic de Sitter phase, as our own. As a result the number of
e-foldings is not larger than 65/85 depending on the type of matter
considered. 

Here we reconsider the problem from the point of view of the entropy
content of the Universe, and considering the correspondence between the
Friedmann equation and Cardy formula in Brane Universes, as discussed by
us \cite{binabdsu}.

The main points in our argument are the following. First we assume a
FRW closed universe with a positive cosmological constant which does not
recollapse, as implied by some recent observations. Such an assumption is
crucial to our argument. Further on, we assume that there is an upper
bound for entropy in the Universe. 
Such an entropy is obtained from the bulk black hole in the sense of
holography and considered to have a bound on its storage to prevent the
collapse of the universe.
These hypothesis are sufficient conditions for us to arrive at the
result. They are nevertheless not necessary, but we think them to be quite
natural. Later on, we extend the result for a Randall-Sundrum brane world
model. While dynamical details of the AdS/CFT correspondence have been
used in the derivation, it is the bound on the entropy of the Universe
which is the essential ingredient.

The existence of an upper bound for the number of e-foldings before the
end of inflation was also studied recently in \cite{scott} and
\cite{liddle}. However in their investigation the number of e-foldings is
related both to a possible reduction in energy scale during the late
stages of inflation and to the complete cosmological evolution, being
model dependent. The bound has been obtained in some very simple
cosmological settings, while it is still difficult to be obtained in
nonstandard models. Using the entropy bound, the consideration of
physical details connected with the universe evolution can be avoided. We
have obtained the upper bound for the number of e-foldings for a standard
FRW universe as well as non-standard cosmology based on the brane inspired
idea of Randall and Sundrum models.

The starting point is that the scalar factor, in case of the brane
cosmology, is defined by the Darmois-Israel condition
\cite{darmoisisrael}\cite{binabdsu}\cite{eabdbertha}. We consider a 
bulk metric defined by 
\be  
ds^2_5=-fdt^2+f^{-1}da^2+a^2d\Sigma^2_K,
\ee
where $f=k+\frac{a^2}{L^2}-\frac{m}{a^2}$ and  $L$ is the curvature radius
of AdS spacetime. $k$ takes the values $0, -1, +1$ corresponding to 
flat, open and closed geometrics, and $d\Sigma^2_k$ is the corresponding 
metric on the unit three dimensional plane, hyperboloid or sphere. The 
black hole horizon is located at  
\be   
a_H^2=\frac{L^2}{2}(-k +\sqrt{k^2+4m/L^2}).
\ee
The relation between the parameter $m$ and the Arnowitt-Deser-Misner 
(ADM) mass of the five dimensional black hole $M$ is \cite{cejm}
\be 
M=\frac{3\omega_3}{16\pi G_5}m
\ee
where $\omega_3$ is the volume of the unit 3 sphere, $\rm{Vol(S^3)}$, 
and $G_5$ is the Newton constant in the bulk. It is related
to the Newton constant $G_4$ on the brane as $G_5=G_4 L/2$. 

Here, the location and the metric on the boundary are
time dependent. We can choose the brane time such that $
\dot{a}^2=f^2\dot{t}^2 -f$,
in which case the metric on the brane is given by 
\be 
ds^2_4= -d\tau^2+a^2(\tau)d\Sigma^2_3\quad .
\ee

The Conformal Field Theory lives on the brane, which is the boundary of 
the AdS hole. The energy for a CFT on a sphere with radius $a$, of volume
$V=a^3\omega_3$ is given by $E=\frac{L}{a}M$. The total energy $E$ is not
a constant during the cosmological expansion, but decreases like $a^{-1}$. 
This is consistent with the fact that for a CFT energy density we have 
\be\label{rhocft}
\rho_{CFT}=E/V=\frac{3mL}{16\pi G_5 a^4}=\frac{3m}{8\pi G_4 a^4}\quad .
\ee

The entropy of the CFT on the brane is equal to the Bekenstein-Hawking
entropy of the AdS black hole
\cite{wittenidentification}\cite{garrigasasaki}, which is given by the
area of the horizon measured in bulk Planckian units, as given by
\be
S_{CFT}(4D)=S_{BH}(5D)=\frac{V_H}{4G_5}, \quad V_H=a_H^3\omega_3\quad .
\ee
The area of a 3-sphere in AdS equals the volume of the corresponding
spatial section for an observer on the brane. 

The total entropy $S$ is a constant during the cosmological evolution, but
the entropy density of the CFT on the brane is 
\be   
s = \frac SV = \frac{a_H^3\omega_3}{4G_5}\frac 1{a^3\omega_3}
\mathop{=}^{G_5 =\frac {G_4L}2} \frac{a_H^3}{2G_4La^3}\quad
. \label{entropydensity} 
\ee
In the brane world interpretation we have to fulfill matching conditions
for the gravitational fields due to the immersion of the brane into the
bulk (see e.g. \cite{binabdsu},\cite{darmoisisrael},\cite{eabdbertha}).
From the matching conditions we find now the cosmological equations in the
brane are
\be\label{friedmannfromdarmois}  
H^2=-\frac{k}{a^2}+\frac{m}{a^4}-\frac{1-(\sigma/\sigma_c)^2}{L^2}\quad ,
\ee
where $\sigma_c=\frac{3}{8\pi G_5 L}$ is the critical brane tension.
Taking $\sigma=\sigma_c$, (\ref{friedmannfromdarmois}) reduces to the 
Friedmann equation of CFT radiation dominated brane universe without 
cosmological constant discussed in \cite{savoverli}. If $\sigma > 
\sigma_c$ or $\sigma < \sigma_c$, the
brane-world is a de Sitter universe or AdS universe, respectively. Using 
(\ref{rhocft}) the Friedmann equation can be written in the form
\be 
H^2= -\frac k{a^2} + \frac{8\pi G_4}{3}\rho_{CFT} +\frac{\lambda}3\quad ,
\ee
where $\lambda$ is the effective positive cosmological constant in four
dimensions, in agreement with observations. The arguments and formulae
above depend on the holographic properties, which we suppose to be valid in
the theory. 

The relation between the energy density and the entropy, 
$\rho =\frac{9mL}{16\pi^2a_H^3a^4}S $ can be used to rewrite the Friedmann
equation as
\be  
\left({\dot a}\right)^2 + k - \frac{3G_4mLS}{2\pi
a_H^3a^2} -\frac{\lambda}3 a^2=0\quad , \label{mechanicalproblem}
\ee
which corresponds to the movement of a mechanical nonrelativistic particle
in a given potential. This equation is crucial for the developments which
follow, being deeply rooted on (\ref{entropydensity}), which is a direct
consequence of holography and the subsequent construction, as given in
{\it e. g. } \cite{binabdsu}. For a closed universe there is a critical
value for which the solution extends to infinity (no big crunch) which is
\be\label{entropyinequality}   
S < \frac{2\pi a_H^3\lambda a^4}{9G_4mL}\mathop{\rightarrow}^{a\approx
\lambda^{-1/2}} \frac{2\pi a_H^3\lambda^{-1}}{9G_4mL}\quad .
\ee
$\lambda^{-1/2}$ is the size of the de Sitter horizon, which is the box
holding the maximum amount of the entropy. 
The above equation was obtained by considering the potential obtained from
the mechanical problem (\ref{mechanicalproblem}) for a closed universe,
namely $k=1$, which is $V= 1-\frac{3G_4mLS}{2\pi
a_H^3a^2}-\frac{\lambda}{3}a^2$. We want a solution which does not collapse
to zero, but rather develops to infinity. The maximum of the potential
occurs at $S_{max}=\frac{2\pi a_H^3\lambda a^4}{9G_4mL} $ and divides the
collapsing region (for $a$ smaller than the one corresponding to the
maximum) or an expanding region, (for $a$ larger than the one
corresponding to the maximum). For the expanding case the first inequality
in (\ref{entropyinequality}) must be fulfilled.

Now, dividing (\ref{entropydensity}) by (\ref{rhocft}) we get
$s=\frac{4\pi a_H^3}{3mL}a\rho$, thus the entropy in such a universe can be
rewriten as 
\be\label{entropyequalsigmav}   
S= s V = \frac 43 \pi a^3\frac{4\pi a_H^3a}{3mL}\rho\quad 
\ee
at the end of inflation. We take $\rho$ to be the energy density during
inflation, that is, $\rho\sim   \frac 1{8\pi G_4} \Lambda_I$, which for
the scale factor at the exit of inflation leads to the value $ a\approx
\Lambda_I^{-1/2} e^{Ne}$, where $\Lambda_I^{-1/2}$ corresponds to the
apparent horizon during inflation, and we obtain
\be
N_e = \ln a +\frac 12 \ln \Lambda _I\quad .
\ee
Using now (\ref{entropyequalsigmav}) in (\ref{entropyinequality}) we get
\be
a< \left(\Lambda_I\lambda \right)^{-1/4}\quad , \label{scalefactorbound}
\ee
from which we arrive at
\be
N_e < \frac 14 \ln \frac{\Lambda_I}{\lambda} 
\approx 64\quad , \label{firstefoldbound}
\ee
where we used the usual values $\Lambda_I^{1/4}\approx 10^{16}$GeV and
$\lambda^{1/4}\approx 10^{-3}$eV. Note that the bound 
(\ref{scalefactorbound}) is stricter than the previous $a \approx
\lambda^{-1/2}$ used before. The bound (\ref{entropyinequality}) implies a
bound for the scale factor such that  only limited amount of entropy can
store to avoid the big crunch. This is the reason we need
(\ref{scalefactorbound}) in order to get the result (\ref{firstefoldbound}).

For this standard FRW universe, the bound obtained is in agreement with
\cite{scott} and \cite{liddle} as well as with \cite{banksfischler}
for a universe filled with radiation. The de Sitter-closed universe
satisfies the correspondence  between Friedmann equation and Cardy
formula, which is the extension of Verlinde's
argument (see \cite{wangabdallasu}) showing the spirit of holography.
It is doubtful that a similar bound can be obtained along the same lines
for open or flat universes. However, we stress the fact that recent WMAP
analysis favours a closed universe, although this is still a result to be
further confirmed \cite{nature}.

Let us consider now very high energy brane corrections to the Friedmann
equation. From the Darmois-Israel conditions we find 
\be
H^2=-\frac k{a^2} + \frac{8\pi}{3M_4^2}\rho
+\frac{4\pi}{3M_4^2}\frac{\rho^2}{l} +\frac{\lambda}{3}\approx
-\frac k{a^2} +\frac{4\pi}{3M_4^2}\frac{\rho^2}{l} +\frac{\lambda}{3}\quad ,
\ee
where $l$ is the brane tension and in the very high energy limit the $\rho^2$
term dominates. $M_4$ and $\lambda$ are four-dimensional Planck scale and
cosmological constant, respectively. 
Within the high-energy regime, the expansion laws corresponding to matter
and radiation domination are slower than in the standard cosmology
\cite{liddle}. Slower expansion rates lead to a larger value of the number
of e-foldings. However, the full calculation has not been obtained due to
the lack of knowledge of this high-energy regime. Here we study this
problem from the point of view of holography.

The energy density of the CFT and the entropy density are related as follows,
\bear
\rho_{CFT}&=& \frac{3m}{8\pi G_4 a^4}\quad ,\quad s =
\frac{a_H^3}{2G_4La^3} \quad ,\nonumber\\
\rho &=& \frac{9mLS}{16\pi^2a_H^3a^4}\quad ,
\eear
which can be substituted in the Friedmann equation as before, leading to a
bound for the entropy, as well as a bound for the scale factor, as given by
\bear
S^2=  \frac{256\pi^4a_H^6a^8\rho^2}{81m^2L^2}  &<&  
\frac{64\pi^3la_H^6}{243G_4m^2L^2\lambda^3}\nonumber\\
a^8 &<&  \frac{3l}{4\pi G_4\rho^2\lambda^3}\quad ,
\eear

We consider the era when the quadratic energy density is important. The
brane tension is required to be bounded by \cite{maartenswands}
$l<\left(1 {\rm MeV} \right)^4$. Combining the values of $\Lambda_I$ and
$\lambda$ we chose before, a bound for $N_e$ is given by
\be
N_e < 75\quad.
\ee

The number of e-foldings obtained is bigger than the value in standard FRW
cosmology, which is consistent with the argument of \cite{liddle}.

In summary, we have derived the upper limit for the number of e-foldings
based upon the arguments relating Friedmann equation and Cardy
formula. For the standard FRW universe our result is in good agreement
with \cite{scott} and \cite{liddle} and in the radiation dominated case
with \cite{banksfischler}. For the brane inspired cosmology in four
dimensions we obtained a larger bound. Considering such a high energy
context, the expansion laws are slower than in the standard cosmology, and
our result can again be considered to be consistent with the argument in
\cite{liddle}. The interesting point here is that using the holographic
point of view, we can avoid a complicated physics during the universe
evolution and give a reasonable value for the upper bound of the number of
e-foldings. Elsewhere \cite{scott,liddle} the mechanism of
ending inflation and the reheating phase are very important. Therefore in
those discussions there is a strong  model dependence. In the present
description using the holographic description we do not
refer to those sensitive processes. We thus claim that this
discussion is more general.

ACKNOWLEDGEMENT: This work was partically supported by  
Fundac\~ao de Amparo \`a Pesquisa do Estado de
S\~ao Paulo (FAPESP) and Conselho Nacional de Desenvolvimento
Cient\'{\i}fico e Tecnol\'ogico (CNPQ).  B. Wang would  
like to acknowledge the support given by 
NNSF, China, Ministry of Science and
Technology of China under Grant No. NKBRSFG19990754 and Ministry of
Education of China. We would like also to thank S. Nojiri for a
useful correspondence.

\end{document}